\documentclass[twocolumn,showpacs,preprintnumbers,amsmath,amssymb,superscriptaddress,floatfix]{revtex4-1}

\usepackage{graphicx}
\usepackage{graphics}
\usepackage{dcolumn}
\usepackage{bm}
\usepackage[T1]{fontenc}

\begin{document}

\title{Structural analysis of strained LaVO$_3$ thin films}

\author{H. Rotella}\email{nnihr@nus.edu.sg}
\altaffiliation[Present address: ]{NUSNNI-NanoCore, National University of Singapore, Singapore 117411}
\affiliation{Laboratoire CRISMAT, UMR 6508 CNRS, ENSICAEN et Universit\'e de Caen Basse Normandie, 6 Boulevard Mar\'echal Juin, F-14050 Caen, France}

\author{O. Copie}
\altaffiliation[Present address: ]{CEA-Saclay, DSM/IRAMIS/SPEC, F-91191 Gif-sur-Yvette Cedex, France }
\affiliation{Laboratoire CRISMAT, UMR 6508 CNRS, ENSICAEN et Universit\'e de Caen Basse Normandie, 6 Boulevard Mar\'echal Juin, F-14050 Caen, France}

\author{G. Mouillard-St\'eciuk}
\affiliation{Laboratoire CRISMAT, UMR 6508 CNRS, ENSICAEN et Universit\'e de Caen Basse Normandie, 6 Boulevard Mar\'echal Juin, F-14050 Caen, France}

\author{H. Ouerdane}
\altaffiliation[Present address: ]{Laboratoire Interdisciplinaire des Energies de Demain (LIED), UMR 8236 Universit\'e Paris Diderot, CNRS, 5 Rue Thomas Mann, 75013 Paris, France}
\affiliation{Laboratoire CRISMAT, UMR 6508 CNRS, ENSICAEN et Universit\'e de Caen Basse Normandie, 6 Boulevard Mar\'echal Juin, F-14050 Caen, France}

\author{P. Boullay}\email{philippe.boullay@ensicaen.fr}
\affiliation{Laboratoire CRISMAT, UMR 6508 CNRS, ENSICAEN et Universit\'e de Caen Basse Normandie, 6 Boulevard Mar\'echal Juin, F-14050 Caen, France}

\author{P. Roussel}
\affiliation{Laboratoire UCCS, CNRS UMR 8181, ENSCL, Bat C7, BP 90108 F-59652 Villeneuve d'Ascq, France}

\author{M. Morales}
\affiliation{Laboratoire CIMAP, CNRS UMR 6252, ENSICAEN et Universit\'e de Caen Basse Normandie, 6 Boulevard Mar\'echal Juin, F-14050 Caen, France}

\author{A. David}
\affiliation{Laboratoire CRISMAT, UMR 6508 CNRS, ENSICAEN et Universit\'e de Caen Basse Normandie, 6 Boulevard Mar\'echal Juin, F-14050 Caen, France}

\author{A. Pautrat}
\affiliation{Laboratoire CRISMAT, UMR 6508 CNRS, ENSICAEN et Universit\'e de Caen Basse Normandie, 6 Boulevard Mar\'echal Juin, F-14050 Caen, France}

\author{B. Mercey}
\affiliation{Laboratoire CRISMAT, UMR 6508 CNRS, ENSICAEN et Universit\'e de Caen Basse Normandie, 6 Boulevard Mar\'echal Juin, F-14050 Caen, France}

\author{L. Lutterotti}
\affiliation{Dipartimento di Ingegneria Industriale, Univ. di Trento, 9 via Sommarive, 38123 Trento, Italy}

\author{D. Chateigner}
\affiliation{Laboratoire CRISMAT, UMR 6508 CNRS, ENSICAEN et Universit\'e de Caen Basse Normandie, 6 Boulevard Mar\'echal Juin, F-14050 Caen, France}

\author{W. Prellier}
\affiliation{Laboratoire CRISMAT, UMR 6508 CNRS, ENSICAEN et Universit\'e de Caen Basse Normandie, 6 Boulevard Mar\'echal Juin, F-14050 Caen, France}

\date{\today}

\begin{abstract}
While structure refinement is routinely achieved for simple bulk materials, the accurate structural determination still poses challenges for thin films due on the one hand to the small amount of material deposited on the thicker substrate and, on the other hand, to the intricate epitaxial relationships that substantially complicate standard X-ray diffraction analysis. 
Using a combined approach, we analyze the crystal structure of epitaxial LaVO$_3$ thin films grown on (100)-oriented SrTiO$_3$. Transmission electron microscopy study reveals that the thin films are epitaxially grown on SrTiO$_3$ and points to the presence of 90$^{\circ}$ oriented domains. The mapping of the reciprocal space obtained by high resolution X-ray diffraction permits refinement of the lattice parameters. We finally deduce that strain accommodation imposes a monoclinic structure onto the LaVO$_3$ film. The reciprocal space maps are numerically processed and the extracted data computed to refine the atomic positions, which are compared to those obtained using precession electron diffraction tomography. We discuss the obtained results and our methodological approach as a promising thin film structure determination for complex systems.
\end{abstract}

\maketitle
\section{Introduction}
Transition metal oxides (TMOs) form a class of materials that exhibits a broad spectra of functional properties such as, {\it e.g.}, metal-insulator transition, ferroelectricity, superconductivity, and colossal magnetoresistance \cite{smolenskii64,wang03,dijkkamp87,hamet92,imada98,david10}. They originate from the particular electronic and atomic structures of TMOs, which induce high electronic polarizability and strong Coulomb correlations; and unlike conventional semiconductors or metal, there is no dominant mechanism in TMOs that dictates their macroscopic properties: all, amongst the interactions that give rise to strong coupling between lattice, electric charges, spins and orbitals in these systems, compete with comparable strength. Consequently, the ground-state landscape in TMOs exhibits a rich structure of low-energy phases, which may be explored at little energy expense \cite{rondinelli11}. 

Among the oxide compounds, the so-called Mott insulators \cite{hubbard63} are typical systems within which the competition between electron kinetic energy and Coulomb repulsion yields the formation of an energy gap. The crystal chemistry also plays an important role in the system property: as it affects electron transport, it necessarily influences the metal-to-insulator transition. Furthermore, these materials are very sensitive to external constraints such as temperature or hydrostatic pressure. Hence, taking advantage of the substrate-induced biaxial strain, the thin film deposition provides a convenient way to tune the TMOs' properties \cite{wang03,dijkkamp87,david10,hamet92,salvador99}. For instance, octahedral rotations in \emph{AB}O$_3$ compounds having a direct effect on the orbital overlaps through the B-O-B angles, may modify their transport properties. The majority of the perovskite group compounds are distorted derivatives of the parent cubic ($Pm\bar{3}m$ (\#221)) resulting from a combination of the following contributions : (i) tilting of $BO_6$ octahedra; (ii) Jahn-Teller distortion of $BO_6$ octahedra; (iii) cation displacement. Several theoretical studies have already addressed the crucial role of these contributions on electronic properties \cite{rondinelli11}.
As the control of the properties in these systems is extremely important in view of their potential applications, it has thus become crucial to gain a fine understanding of their structure. Many progresses have been made in the synthesis of perovskite thin films and complex heterostructures but since little is known about their actual crystal structures, it remains difficult to quantify, for a thin film, the relationships between its physical properties and its lattice structure. When comes the question of structural analysis of complex systems as epitaxial oxide thin film grown on an isostructural substrate, the accurate structure refinement still remains challenging. 

X-ray diffraction (XRD) is the most widely used non-destructive analytical technique, which reveals relevant informations on a crystal structure. However, in case of a thin film deposited on a substrate, the geometry of the sample and the small diffracting volume strongly reduce the interest of this technique. First, the epitaxy of the film produces a highly textured material which often presents oriented crystallographic domains, resulting in a complex diffracted pattern with the convolution of several crystallographic planes contribution in one peculiar reflection. Second, the isostructural relation between the film and the substrate structure produces two diffracted patterns in the reciprocal space which are very close to each other. The amount of substrate material being 5 orders of magnitude larger than that of the film, the diffracted beam coming from the substrate is much more intense than the film signal. If the structures are close, the deconvolution of the two signals becomes difficult. Third, as the small diffracting volume of the film produces a weak diffracted beam, this may result in the apparent extinction of the weak reflections in the diffracted pattern which should normally be present because of the peculiar film structure. In addition to those sample constraints, the experimental set-up itself produces limitations in the data acquisition: the system, film and substrate, only allows the reflection configuration. In this specific configuration, the experimental set-up produces shadow zones in the diffracted pattern, resulting in inaccessible reflections.\cite{moram09}

One solution to overcome these constraints is to use a combination of several techniques focusing at specific points. The combination of transmission electron microscopy (TEM) and XRD analyses results in a first knowledge of the film structure and microstructure. Second, a finer structure analysis can be operated by separately focusing on the epitaxial relations between the film and the substrate and on the refinement of the lattice parameters of the film. Note that even the determination of the atomic positions can be made separately by focusing on a particular species of atoms. In the case of oxide perovskite compounds, it as been proved that the displacements of the oxygen atoms from their ideal positions produce specific reflections in the diffracted pattern \cite{glazer72, glazer75}. Using XRD, partial structural studies focused on the determination of the amplitude of octahedral tiltings in thin films were achieved \cite{may10,may11,rotella12,vailionis11,lu13}. 
Dealing with the structure determination of unknown phases deposited in the form of thin films, precession electron diffraction (PED) \cite{vincent94} has proved to be one valuable technique \cite{boullay09,simon13} not limited by the size of the probe nor the small volume of diffracting material. This technique was thus retained as a good tool to investigate the structure of our LaVO$_3$ (LVO) thin films grown on (001)-oriented SrTiO$_3$ (STO) substrates. In its bulk form, LVO crystallizes at room temperature in an orthorhombic structure ($Pnma$ (\#62)) with the following lattice parameters, $a_o$=5.5529(2) \AA , $b_o$=7.8447(3) \AA\ and $c_o$=5.5529(3) \AA\ \cite{bordet93}. The LVO structure presents tilting of $BO_6$ octahedra and La displacements which makes it a derivate of the parent cubic structure ($Pm\bar{3}m$). However in the case of LVO these distortions are relatively small and the lattice parameters can be related to a pseudocubic structure according to the following equation : $a_p\simeq a_o/\sqrt{2}\simeq b_o/2\simeq c_o/\sqrt{2}\simeq 3.9251(1)$ \AA . The mismatch between $a_{\rm{p_{LVO}}}$ and $a_{\rm{STO}}$ is about 0.5\% indicating a film compressive stress. 
In this paper, we present an in-depth study of the structure of LVO thin films by combining XRD and precession electron diffraction tomography (PEDT) \cite{kolb07}. Then, we discuss the obtained results by our approach as a promising thin film structure determination for complex systems.

\section {Experimental}

Epitaxial LaVO$_3$ (LVO) thin films were grown by pulsed laser deposition technique on (001)-oriented SrTiO$_3$ (STO) substrate (cubic $a$=3.905 \AA ). To grow the films, a KrF laser ($\lambda$=248 $nm$) with a repetition rate of 2 Hz and a fluence of $\simeq$ 2J/cm$^2$ was focused onto a LaVO$_4$ polycrystalline target. The substrate was kept at 700$^{\circ}$C under a dynamic vacuum near $10^{-5}$ mbar. The distance between the target and the substrate is 8.5 cm.\cite{hotta06,sheets09}

The sample used for transmission electron microscopy (TEM) analysis was prepared in cross-section using a JEOL ion slicer. After a hand polishing with a series of grinding paper, the cross-section was finished by using an Ar ion beam to decrease the thickness down to 100 nm. High resolution electron microscopy (HREM) images were obtained using a FEI Tecnai G2 30 (LaB6 cathode) microscope. 
Precession Electron Diffraction tomography data were recorded on a JEOL 2010 (LaB6 cathode) microscope equipped with a Nanomegas DigiStar precession module and an upper-mounted Gatan Orius CCD camera. 64 PED patterns were recorded in the tilt range -33 to +30 degrees with a precession angle of 1.5 degree.
The data analysis and reduction were performed using the programs PETS\cite{palatinus11} and Jana2006\cite{jana} following a procedure similar to the one described in ref.\cite{boullay13}.


The reciprocal space maps (RSM) were acquired using a high resolution 7-circles Rigaku SmartLab diffractometer having a copper rotating anode ($\lambda = 1.54056$ \AA ) and a 1D detector of 2$^{\circ}$. This apparatus presents weak wavelength dispersion ($\Delta \lambda/\lambda=3.8\times10^{-4}$) and weak beam divergence ($\Delta \omega =32'$). 
After a classical optimization of the diffractometer angles on the substrate, for intensity measurements of the film, the angles were adjusted to maximize the film intensity.
We scanned a total of 26 RSMs either in coplanar or non-coplanar configurations.

\begin{figure}[ht]
\centering
\includegraphics[scale=0.3]{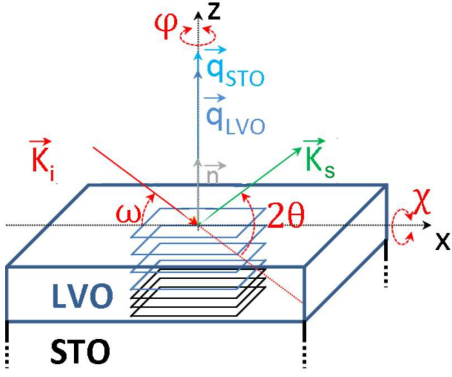}
\caption{Schematic representation of the measurement configuration and definition of the sample reference coordinate systems. $\vec{k_i}$ and $\vec{k_s}$ represent, respectively, the incident and scattered beam vectors.}
\label{ref}
\end{figure}

The sample reference frame (Fig.\ref{ref}) used in the RSM corresponds to \textbf{$\overrightarrow{x}$} and \textbf{$\overrightarrow{y}$} axes aligned with the main edges of the sample, and \textbf{$\overrightarrow{z}$} perpendicular to the sample surface. The diffractometer rotation axes are the conventional $\chi$ (tilt), $\omega$ (incident) and $\phi$ (azimuth) angles, with the incident X-ray beams at $\omega$ from the sample plane at $\chi$=0, and its projection on the sample plane aligned with the \textbf{$\overrightarrow{x}$} axis at $\phi$=0. The fourth circle of the diffractometer corresponds to the Bragg angle, 2$\theta$, given by the detector position.

When rotating the sample around $\chi$, $\omega$ and $\phi$, the beam penetrates the sample in different ways, and, as shown in the following equations, a correction of the diffracted intensity at each ($\chi$, $\omega$, $\phi$, $\theta$) measured point becomes necessary \cite{chat2010} before analysis: 

\begin{equation} \label{corr}
I_{\rm corr}=\frac{I_{\rm mes}}{A_{\chi}^{\rm film}}
\end{equation}
where $A_{\chi}^{\rm film}$ is the correction factor:
\begin{equation} \label{A}
{A_{\chi}^{\rm film}}=\frac{2}{\sin 2\theta M(\omega ,2\theta )} \frac{1-\exp \left[\frac{-\mu TM(\omega ,2\theta )}{\cos \chi }\right]}{1-\exp \left[\frac{-2\mu T}{\sin 2\theta\cos\chi}\right]} 
\end{equation}
with T the effective thickness of the sample and 
\begin{equation} \label{M}
M(\omega ,2\theta )= \frac{1}{\sin 2\theta}+\frac{1}{\sin (2\omega -2\theta )}.
\end{equation}

\begin{figure}[ht]
\centering
\includegraphics[scale=0.35]{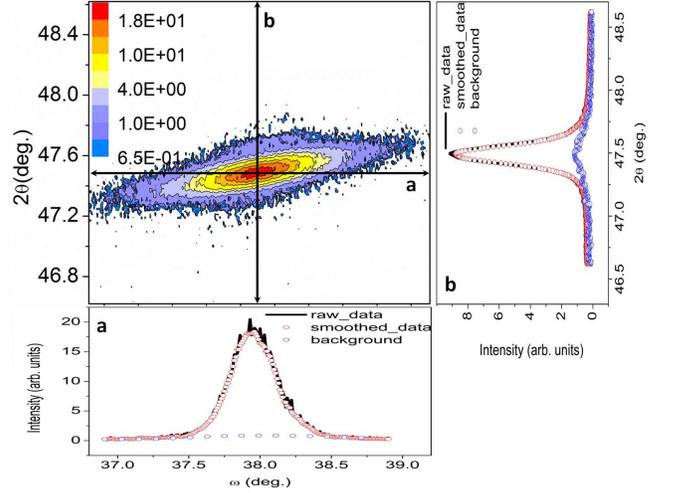}
\caption{(212) Reciprocal space map of LVO. The raw and smoothed data, with the associated background, for a $\omega$ scan at a given 2$\theta$ value (represented by the \textbf{a} arrow ) is plotted in (\textbf{a}). The distribution of the intensity of the smoothed data and the background along 2$\theta$ represented by the (\textbf{b}) arrow and plotted in (\textbf{b}).}
\label{212}
\end{figure}

Once this correction is applied, we process the corrected RSM. Each RSM is made of a series of 1D diffractograms with 2$^{\circ}$ intervals in $\omega$ for one fixed 2$\theta$. To calculate the integrated intensity of each RSM, we developed a program that fits each ($\omega$,2$\theta$) scan (Fig.\ref{212}).

\section {Results and Discussion}

\begin{figure}[ht]
\centering
\includegraphics[scale=0.47]{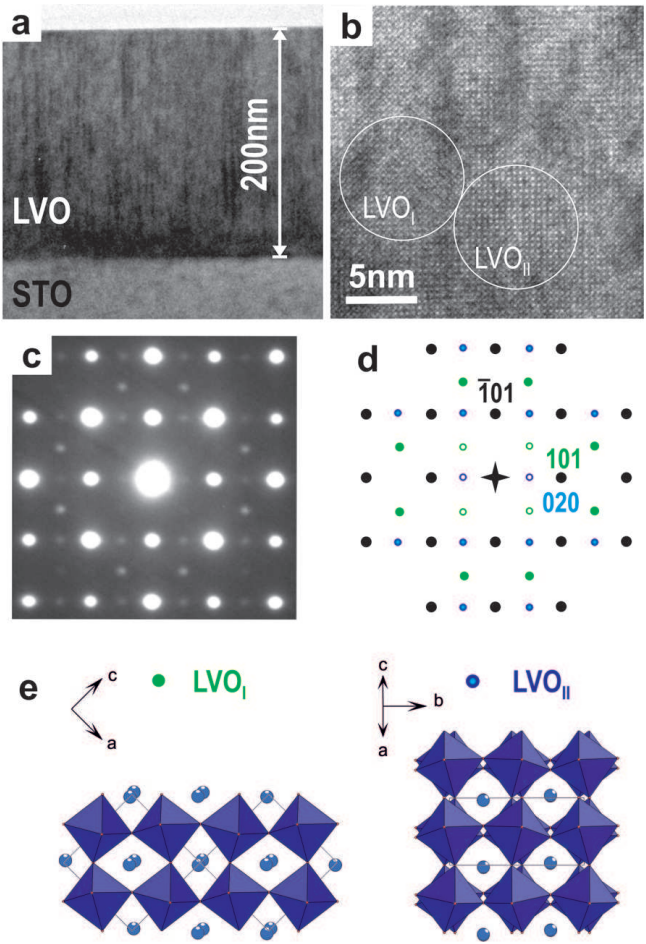}
\caption{\textbf{(a)} TEM bright field view and \textbf{(b)} HREM image of a LVO thin film grown on a (001)-oriented SrTiO$_3$ substrate. \textbf{(c)} SAED patterns obtained from an area corresponding to the whole film thickness. \textbf{(d)} schematic indexing of the SAED patterns with the contribution of 90$^{\circ}$-oriented domains (SG: Pnma) represented in \textbf{(e)}. Black spots correspond to reflections common to both orientations (perovskite subcell). Reflections specifically related to LVO$_{\rm{I}}$ and LVO$_{\rm{II}}$ are indicated, respectively, in green and blue.}
\label{micro}
\end{figure}

Preliminary investigations were made by TEM on a 200 nm-thick film. The TEM bright field image obtained along one $<$100$>_{\rm{STO}}$ direction (Fig.\ref{micro}a) presents a high crystalline quality with a perfect epitaxy of the film on the substrate. In the whole thickness, the film is made off the imbrication of oriented nanosized domains as evidenced by HREM imaging (Fig.\ref{micro}b) and selected area electron diffraction (SAED) patterns (Fig.\ref{micro}c). In this pattern, the most intense reflections correspond to the perovskite subcell common to both STO and LVO phases and are indicated by black dots in the schematic Fig.\ref{micro}d. 
The weak reflections denote the presence of a superstructure as referred to the Pm$\bar{3}$m prototype perovskite. They can be indexed if one considers that the LVO film presents a distorted perovskite structure involving rotation of VO$_6$ octahedra consistent with the ones existing in bulk LVO \cite{bordet93}. In this case, considering the existence of 90$^\circ$ oriented domains, the weak reflections can be separated in two subsets associated, respectively, to [010] and [101] zone axes patterns of a Pnma structure having cell parameters a$_{\rm{p}}\sqrt{2}\times$2a$_{\rm{p}}\times$a$_{\rm{p}}\sqrt{2}$.
In Fig.\ref{micro}e, these two orientations correspond to the case where the b-axis of the Pnma LVO structure is parallel to the substrate plane but differs by an in-plane rotation of 90$^\circ$ around $\vec{z}$. Considering its Pnma bulk form, the epitaxial relationship for LVO can be written as (101)LVO$\parallel$(001)STO with for LVO$_{\rm{I}}$: [010]LVO$\parallel$[100]STO and for LVO$_{\rm{II}}$: [010]LVO$\parallel$[010]STO. Interestingly, the microstructure of LVO deposited in this 200nm thick film offers strong similarities with the one observed for thin LVO layers in LVO/SrVO$_3$ heterostructures.\cite{boullay11}

\begin{figure}[ht]
\centering
\includegraphics[scale=0.35]{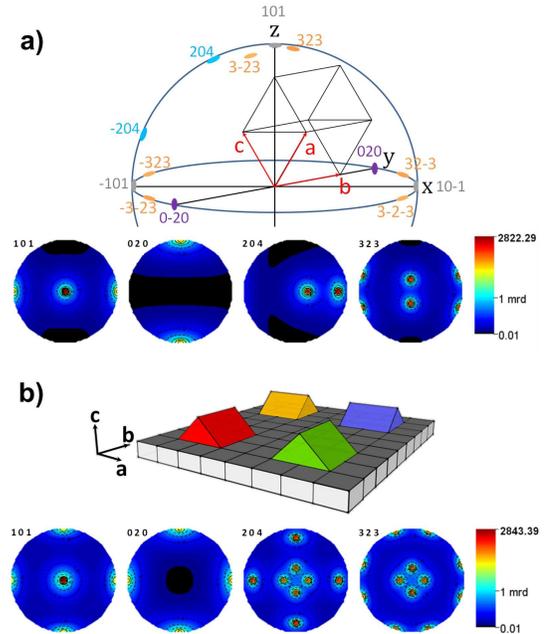}
\caption{a) (upper) schematic representation of the pole sphere for a perfect epitaxial thin film of LVO. (lower) corresponding {101}, {103} and {204} pole figures.
b) (upper) schematic representation of the four 90$^{\circ}$ oriented domains. (lower) corresponding {101}, {020}, {204} and {323} directions. Each component is represented as a 10$^{\circ}$ HWHM Gaussian contribution.}
\label{PF-sim}
\end{figure}

A perfect epitaxial film, without any growth defects nor orientation domains {\it i.e.} a single crystal would behave like a perfectly textured sample with one orientation component. However, most of the epitaxial thin films are assemblies of several crystallites with different orientations. In the case of our LVO thin film, according to the TEM analysis, four 90$^{\circ}$ oriented domains are present in the sample {\it i.e.} the two represented in Fig.\ref{micro}e plus their equivalents by a 180$^\circ$ in plane rotation around $\vec{z}$. In order to get a better view on how these oriented domains shall affect the RSM measured by XRD, the simulation of the expected pole figures, represented in equal area projections \cite{chat2010}, were performed with the MAUD software \cite{maud} considering the bulk LVO structure \cite{bordet93}. 
For simplicity, we focus only on the $101$, $020$, $204$ and $323$ reflections. First, considering that the film is fully textured, {\it i.e.} with only one orientation component, 
we simulate the four pole figures (Fig.\ref{PF-sim}a). Each pole, as well as its multiplicity, is clearly observed independently from the others. 
Second, the four 90$^{\circ}$ oriented domains of the film are considered for the simulations (Fig.\ref{PF-sim}b). 
The convolution of several poles appear at the same location in the pole sphere. At the equator of the pole sphere, the $101$ and the $020$ equivalents are mixed; so do the $204$ and $323$ equivalents near the north pole. Following the TEM observations, these simulations indicate that the presence of 90$^{\circ}$-oriented domains implies that some reflections represent the contributions of several crystallographic planes.

\begin{figure}[ht]
\centering
\includegraphics[scale=0.44]{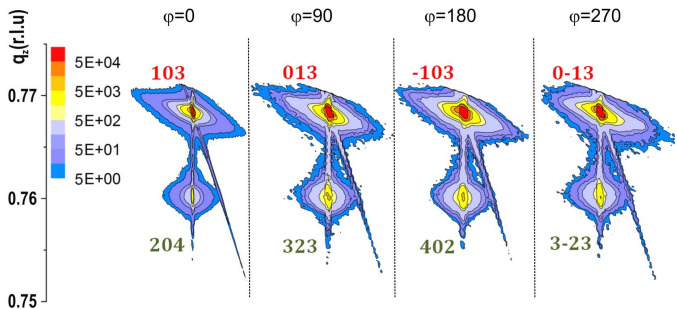}
\caption{Reciprocal space maps along the $<$103$>^*$ SrTiO$_3$ and [204]$^*$, [402]$^*$, [323]$^*$, [3-23]$^*$ of LVO directions. The horizontal axis is $q_x$ for each of the 4 RSMs.}
\label{204}
\end{figure}

To further investigate the structure of LVO in a thin film form, we performed high resolution XRD measurements on a 100 nm-thick film. First, we focused on the characterization of the epitaxial relationships between the film and the substrate examining the four asymmetric reflections (204), (402), (323) and (3-23) shown in Fig.\ref{204}. These are represented in the reciprocal lattice units q$_x$ = $\frac{2 \pi}{\lambda}\left[\cos(2\theta - \omega) - \cos\omega\right]$, q$_z$ = $\frac{2 \pi}{\lambda}\left[\sin(2\theta - \omega) + \sin\omega\right]$, where $\lambda$ is the wavelength and $\theta$ and $\omega$ are the angles described (Fig.\ref{ref}).
This representation is valuable for these reflections because the measurements are performed in the coplanar configuration, {\it i.e.} the angle $\chi$ in the Fig.\ref{ref} is kept equal to zero. To reach these reflections, the angle $\omega$ is defined by the relation $\omega = \theta -\chi$.

The RSM (Fig.\ref{204}) shows that the LVO film is coherently and fully strained on the substrate, {\it i.e.} we observe a perfect vertical alignment between both families of reflections from film and substrate. As expected from the pole figure simulations (Fig.\ref{PF-sim}b), the horizontal alignment of these four reflections reveals that they are equivalent due to the presence of oriented domains in the film. We can conclude from this observation that the $a_p$ and $b_p$ lattice parameters of LVO thin film are qualitatively very close. Also, the relative fractions of the four 90$^{\circ}$ oriented domains are equal, with equivalent measured intensity for the four groups of reflections. This is in agreement with the TEM observations described above.

To perform cell parameters refinement, we recorded all the accessible reflections allowed by the diffractometer set-up (Table \ref{reflexion}). Unfortunately, we observe a huge distortion of the beam imprint with sample tilts (when $\chi$ is nonzero). This effect yields an important error on the absolute value in 2$\theta$ of the RSM ellipse center. This can be corrected by using an analyzer crystal in front of the detector but at the cost of an intensity decreased by 2 orders of magnitude. Consequently, several reflections coming from the film could not be reached. This explains why certain reflections in table \ref{reflexion} are not associated with a 2$\theta$ position. 

\begin{table}[ht]

\renewcommand{\arraystretch}{1.25}

\centering
\begin{tabular}{c c c c c c c c c}
\hline\hline
$h$ $k$ $l$ & & $2\theta$($^{\circ}$) & I$_{int}$ & \hspace{0.5cm} & $h$ $k$ $l$ & & $2\theta$($^{\circ}$) & I$_{int}$  \\ \hline
1 0 1 & & 22.52(1)  &  118   & & 0 3 3 & &           &  0.012\\ 
1 0 2 & &           &  0.016 & & 3 3 1 & & 63.82(1)  &  0.56\\
2 1 1 & &           &  0.610 & & 4 0 0 & &           &  60\\
2 2 1 & &           &  0.027 & &   & &    &   \\ 
2 0 2 & & 45.97(1)  &  500   & & 4 1 0 & &           &  0.67\\ 
0 4 0 & & 46.47(1)  &        & & 4 0 1 & & 69.59(1)  &  0.011\\
2 1 2 & & 47.48(1)  &  1.65  & & 4 1 1 & & 70.77(1)  &  0.046\\
2 3 1 & &           &  0.14  & & 3 0 3 & & 71.67(1)  &  2.5\\
1 3 2 & &           &  0.14  & & 3 1 3 & & 72.83(1)  &  0.93\\
1 0 3 & &           &  27    & & 4 2 1 & & 74.32(1)  &  \\  
3 1 1 & & 53.34(1)  &  0.77  & & 2 0 4 & & 76.31(1)  &  78\\
1 2 3 & & 57.42(1)  &  151   & & 3 2 3 & & 76.32(1)  &  77\\
2 0 3 & & 59.70(1)  &  0.008 & & 3 $\bar{2}$ 3 & & 76.31(1)  &  78\\
3 1 2 & & 61.00(1)  &  0.016 & & 4 0 2 & & 76.31(1)  &  79\\
\hline\hline
\end{tabular}
\caption{2$\theta$ positions and integrated intensities of the measured reflections. Neither the 2$\theta$ positions of the non-coplanar reflection measured without analyzer, nor the integrated intensity of both 040 and 421 (different set-up configuration for the 040 and too much distortion on the 421), are reported.}
\label{reflexion}
\end{table}

From the list of absolute 2$\theta$ positions, we refined the cell parameters of the LVO thin film using the CELREF software.\cite{celref} We showed in a previous work that a LVO thin film grown on SrTiO$_3$ has a distorted structure towards monoclinic symmetry.\cite{rotella12} According to group-subgroup relation, the symmetry lowering form orthorhombic to monoclinic with the appearance of a $\beta$ angle would lead to the space group $P2_1/m$\cite{howard98} (most symmetric choice).
Thus, the refinement procedure was done using $P2_1/m$ leading to the lattice parameters $a=5.554(3)$ \AA , $b=7.810(4)$ \AA ~and $c=5.555(5)$ \AA\ with the monoclinic angle $\beta=89.45(9)^{\circ}$ (table \ref{bilan1}).
The $b$ parameter verifies the relation $b=7.810(4)$ \AA $=2\times3.905$ \AA~, confirming that the film is fully strained. This quantitative result is consistent with previous qualitative observations.\cite{rotella12} Comparing the bulk values with the refined ones, the $a$ and $c$ parameters remain equal to those of the bulk within our experimental accuracy. But in order to accommodate the substrate strain along [$10\bar{1}$] and [$\bar{1}01$] directions, the angle $\beta$ becomes smaller than 90$^{\circ}$, whereas the $b$ parameter changes significantly.
The strain is evaluated to be $\epsilon_2=(b_B-b)/b_B=0.5\%$, where $b_B$ is the bulk parameter. 

\begin{table}[ht]

\renewcommand{\arraystretch}{1.25}

\centering
\begin{tabular}{c c c}
\hline\hline
         & Bordet \emph{et al.}\cite{bordet93}     & this work     \\ \hline
a        & 5.55548(4) & {5.554(3)}  \\ 
b        & 7.84868(6) & {7.810(4)}    \\
c        & 5.55349(5) & {5.555(5)}    \\
$\alpha$ & 90.0$^{\circ}$      & {90.0$^{\circ}$}  \\    
$\beta$  & 90.0$^{\circ}$      & {89.45(9)$^{\circ}$} \\
$\gamma$ & 90.0$^{\circ}$      & {90.0$^{\circ}$}  \\
\hline\hline
\end{tabular}
\caption{Cell parameters refined for our LVO thin film deposited on STO.}
\label{bilan1}
\end{table}

Let now consider the possibility to refine the atomic positions using the intensities integrated from the RSMs. Some questions might occur regarding the intensities recorded in a non-coplanar configuration ($\chi\neq0$) for which some distortion of the diffracted beam is observed. In order to retain as many reflections as possible, we assume that the integrated intensity is not influenced by the distortion of the beam when tilting the sample provided the beam imprint stick within the sample surface. Thus, the RSM in coplanar and non-coplanar configurations are investigated to get the maximum number of reflections. Nonetheless, in order to keep the incident intensity on the sample as constant as possible independently of the orientation of the sample, reflections with small $\omega$ and high $\chi$ values were not measured. Likewise, in our integration process, the errors on the measured reflections are not known and in the following structure refinement we consider a unit weight ponderation scheme to account equitably for all reflections, even the weakest.

\begin{table*}[ht]

\renewcommand{\arraystretch}{1.25}

\begin{center}

\begin{tabular}{c | c c c | c c c | c c c c }
\hline\hline
& \multicolumn{3}{c}{ bulk reference \cite{bordet93}} & \multicolumn{3}{c}{thin film (XRD)} & \multicolumn{4}{c}{thin film (PEDT)} \\ \hline
atom & x & y & z  & x & y & z  & x & y & z  & Uiso(\AA$^2$) \\ \hline
La1 & 0.0295(4)  & 0.25    & 0.9951(8)      &  0.007(2) & 0.25 & 0.001(3)   &  0.0127(6) & 0.25 & 0.9965(8) & 0.018(2) \\
V1  & 0.5       & 0     & 0                 &  0.5       & 0     & 0        &    0.5       & 0     & 0    &   0.023(3)     \\ 
O1  & 0.4880(6) & 0.25    & 0.0707(10)      & 0.503(15) & 0.25 & 0.03(2)   &  0.499(4) &  0.25 & 0.052(4) &   0.021(6) \\
O2  & 0.2831(6)   & 0.0387(4) & 0.7168(6)   & 0.262(14) & 0.014(12) & 0.742(15)  & 0.272(3) & 0.039(6) & 0.730(3) & 0.027(5) \\
\hline\hline
\end{tabular}

\end{center}

\caption{ Results of the refinement of LVO thin films using XRD and PEDT data. The structure of the bulk LVO is given as a reference. The refinement were done in the space group Pnma with the cell parameters a=5.554\AA, b=7.810\AA\ and c=5.555\AA. In the XRD refinement, isotropic thermal displacement parameters (Uiso) were refined to an overall value of 0.065\AA$^2$ for all the atomic positions. The reliability factors Robs/wRobs are 9.4/8.1 and 18.2/20.9 , respectively, for the XRD and PEDT refinement.
}
\label{atom}
\end{table*}

\begin{figure}[ht]
\centering
\includegraphics[scale=0.40]{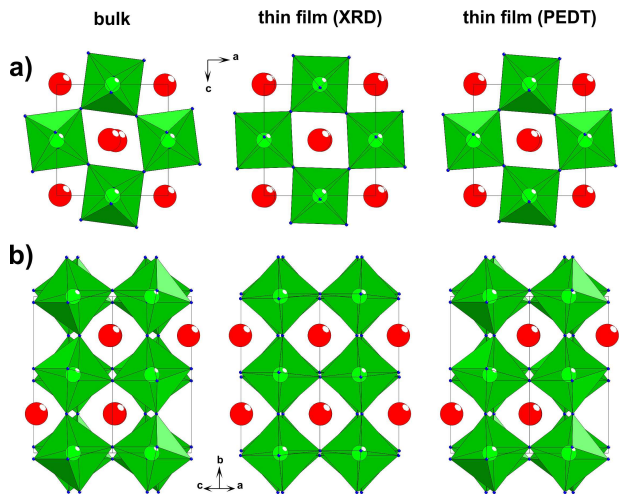}
\caption{Schematic representation of the structure of bulk LVO (room temperature)\cite{bordet93} and LVO thin films as obtained from XRD and PEDT refinements. a) [010] projection b) [101] projection.}
\label{struct}
\end{figure}

While a close analysis of the RSM obtained by XRD evidences a monoclinic distortion of the LVO lattice, the indexing of the 26 measured reflections follows the rules 0kl: k+l=2n and hk0: h=2n compatible with the orthorhombic space group Pnma. 
In the following we thus consider that the LVO thin film structure deviates only slightly to an orthorhombic Pnma structure, leading to a number of 7 atomic parameters to refine (see Table \ref{atom}). This number alone is already too high considering that 10 observed reflections per parameter are usually recommended, without considering the scale factor and the atomic displacement parameters (ADP). Nevertheless, considering a global and isotropic value for the ADP and adding the orientation variants with an equivalent volume fraction, the atomic positions plus the scale factor were refined with JANA2006 \cite{jana}. In the refinement process, to maintain a reasonable geometry of the O$_6$-octahedra, we imposed soft constraints on O-O distances to keep them in the same average value as the ones observed in bulk LVO:2.8\AA $\pm$0.1\AA\cite{bordet93}. Considering this, we obtained the atomic positions indicated in Table \ref{atom} with reliability factors (Robs and wRobs) around 9\%. 

In contrast to XRD, the PEDT data collection allowed to access a much larger number of reflections but with the drawback of being a destructive method and the necessity to prepare a cross-section of the film. Hence, about 3000 reflections were measured leading to 160 unique reflections (averaged from 1750 reflections observed with I>3$\sigma$(I)). 
The structure was refined using the Pnma space group without any O-O distance constraints (Table \ref{atom}). The PED data are biased by dynamical scattering effects still present even using the precession method and the values of the reliability factors (Robs and wRobs around 19\%) are typical of the ones obtained for a refinement against PED data considering the kinematical approximation.

For the various reasons detailled previously, both refinements give reliability factors higher than the standards usually required in crystallography. Correlatively, the obtained atomic positions (Table \ref{atom}) and interatomic distances (Table \ref{dist}) have to be taken with caution especially for the refinement against XRD where the uncertainties on atomic positions and distances are comparatively large. 
When comparing the three structures represented in Fig.\ref{struct}, the octahedral tilting directions and amplitudes look similar between the bulk and the PEDT refinement but differs in amplitude with the XRD refinement. This last refinement clearly lacks of sensitivity regarding oxygen atomic positions which are not accurate enough to draw valuable conclusion. 
The most noticeable difference between the bulk and the thin films structure actually lies on the La atomic position. In the bulk reference, the La is strongly displaced from the 0$\frac{1}{4}$0 position and, with LaO-O distances ranging from 2.42\AA\ and 3.27\AA\cite{bordet93}, the oxygen cuboctahedron surrounding La is strongly distorted with 8 first neighbors at an average distance of 2.60\AA\ and 4 second nearest neighbors at 3.18\AA. In both XRD and PEDT refinements performed on our film, the La environment is significantly less distorted than in bulk reference (Table \ref{dist}) though the PEDT refinement issues with distortions closer to bulk ones. Comparing the structures using the COMPSTRU tool \cite{compstru}, the La atomic position differs between bulk and thin film by a value of 0.127\AA\ and 0.094\AA\ for the XRD and PEDT refinements, respectively. With only a difference of 0.037\AA\ between these two refinements, the La atomic position in the thin film appears significantly different from the bulk. Note that in both XRD and PEDT refinements, the La atomic position is robust compared to the oxygen ones. In our case, this reduction of the La displacement compared to the bulk reference can be regarded as a signature of the substrate induced strain amounting up to stresses of 3.5 GPa \cite{rotella12}.

\begin{table}[ht]

\renewcommand{\arraystretch}{1.25}

\begin{center}

\begin{tabular}{c | c | c }
\hline\hline
& thin film (XRD) & thin film (PEDT)         \\ \hline
V1-O1 (x2) (\AA) &  1.960(9)  &  1.973(3)        \\
V1-O2 (x2) (\AA) &  1.97(9)    &  1.989(19)       \\ 
V1-O2 (x2) (\AA) &  1.97(9)    &  2.000(19)       \\
 Average   (\AA) &  1.97      &    1.99          \\
 & & \\
   min La1-O   (\AA)   &   2.39(8)   &   2.47(4) \\
   max La1-O    (\AA)  &   3.19(8)   &   3.15(4) \\
 La1-O 12        (\AA) &   2.77  &  2.78  \\
 La1-O first 8  (\AA) &   2.71  &  2.65  \\
 La1-O last 4   (\AA) &   2.89  &  3.06  \\
\hline\hline
\end{tabular}

\end{center}

\caption{Selected interatomic distances obtained from the refinement of LVO thin films using XRD and PEDT data. Note that the O-O distances have been constrained to keep an average value of 2.8\AA $\pm$0.1\AA\ in the XRD refinement.
}
\label{dist}
\end{table}

\section{\label{Conclu} Conclusion}

Attempting to accurately determine the structure of an epitaxial thin film is a challenging task even for a "simple" perovskite where finding a structural model is not an issue. In the case of our LVO thin film, we concentrate mainly on the use of diffraction techniques to achieve this goal. From laboratory XRD analysis, we succeed to get a deep insight in the epitaxial relationships and refined accurately the lattice parameters revealing a subtle monoclinic distortion induced by the film compressive stress. The feasibility of the next, decisive, step of refining the structure of an epitaxial thin film against laboratory XRD data is demonstrated despite uncertainties on the oxygen atomic positions. 
In order to get better accuracy, one should increase significantly the number of reflections accessible in the RSM. Certainly one would greatly benefit from the use of synchrotron X-ray diffraction. A higher incoming flux can allow to properly observe the weak reflections in the diffracted pattern of the film. 
Also the acquisition of the RSM using a 2D detector with a dedicated intensity integration software would be a considerable implement in the reliability of the acquired integrated intensity.
Such an experiment would still require a consequent amount of time and an easy access to synchrotron beamline. 
To this respect, Precession Electron Diffraction Tomography is complementary to XRD as being one technique permitting the rapid acquisition of numerous reflections from a thin film cross-section or plane view. Using PEDT data, the structure refinement could be achieved with less uncertainties on atomic positions. In a near future the treatment of intensities using the dynamical theory of electron scattering\cite{palat13} shall allow to achieve more reliable structure refinements. 
In a broader perspective, both techniques can be applied to precisely characterize the structure of various oxide films and correlates it with the electronic properties.

\begin{acknowledgments}
We acknowledge support of the French Agence Nationale de la Recherche (ANR), through the program ``Investissements d'Avenir''(ANR-10-LABX-09-01), LabEx EMC$^3$ and C'NANO project. The authors thank L. Gouleuf for the cross-section preparation and J. Lecourt for target preparation. WP and HR thank Dr. O. Perez (CRISMAT) and Prof. S.~J. May (Drexel Univ.) for illuminating discussions and their continuous encouragements during this work. DC and LL acknowlege the Conseil Régional de Basse-Normandie and FEDER for LL's Chair of Excellence financial. 
\end{acknowledgments}

\end{document}